# Electron transport properties of heterogeneous interfaces in solid electrolyte interphase on lithium metal anodes

Xiangyi Zhou[1]*, Rongzhi Gao[1], Ziyang Hu[1,2]*, Weijun Zhou[3], YanHo Kwok[3], GuanHua Chen[1,2]*

[1]Department of Chemistry, The University of Hong Kong, Pokfulam Road, Hong Kong SAR, China

[2]Hong Kong Quantum AI Lab Limited, Pak Shek Kok, Hong Kong SAR, China

[3]QuantumFabless Limited, Pak Shek Kok, Hong Kong SAR, China

*Corresponding authors. Email: XZ: xy.zhou2025@gmail.com, ZH: hzy@yangtze.hku.hk, and GHC: ghc@everest.hku.hk,

ORCID: XZ: 0009-0002-8311-0768; RG: 0000-0001-7520-0347; ZH: 0000-0002-7693-5457; WJZ: 0000-0002-4328-3704; YHK: 0000-0003-1325-4416; GHC: 0000-0001-5015-0902

# Abstract

In rechargeable batteries, electron transport properties of inorganics in the solid-electrolyte interphase (SEI) critically determine the safety, lifespan and capacity loss of batteries. However, the electron transport properties of heterogeneous interfaces among different solid inorganics in SEI have not been studied experimentally or theoretically yet, although such heterogeneous interfaces exist inevitably. Here, by employing non-equilibrium Green's function (NEGF) method, we theoretically evaluated the atomic-scale electron transport properties under bias voltage for LiF/Li$_2$O interfaces and single-component layers of them, since LiF and Li$_2$O are common stable inorganics in the SEI. We reveal that heterogeneous interfaces orthogonal to the external electric-field direction greatly impede electron transport in SEI, whereas heterogeneous parallel-orientated interfaces enhance it. Structural disorders induced by densely distributed interfaces can severely interfere with electron transport. For each component, single-crystal LiF is highly effective to block electron transport, with a critical thickness of 2.9 nm, much smaller than that of Li$_2$O (19.0 nm). This study sheds a new light into direct and quantitative understanding of the electron transport properties of heterogeneous interfaces in SEI, which holds promise for the advancement of a new generation of high-performance batteries.

# Introduction

In rechargeable batteries with solid or liquid electrolyte, the electrolyte is positioned between the anode and the cathode to facilitate ion transport[1]. The Solid-Electrolyte Interphase (SEI) serves as a critical interface layer between the electrolyte and the anode[1-4]. The characteristics of the SEI layer determine many key properties of rechargeable batteries[5-8]. The SEI can form naturally through self-limiting electrochemical decomposition of the electrolyte, or be introduced during the manufacturing process (known as artificial SEI)[9,10]. Naturally formed SEIs are multi-component, typically with inorganics near the electrode and organics near the electrolyte[11,12]. Among them, LiF and $Li_2O$ are two commonly stable inorganics in SEI[13-15]. Therefore, heterogeneous interfaces between different components in SEI inevitably exist and play a crucial role in determining the properties of the SEI and batteries[14,15]. However, the electron transport properties of heterogeneous interfaces among different solid inorganics in SEI are completely unknown to the best of our knowledge, despite of their importance in determining the safety and lifespan of batteries[16,17].

Electron transport across the SEI is one of the driving forces behind self-limiting electrolyte decomposition and dendrite formation[3]. An ideal SEI should have suitable thickness and be crack-free. Such an ideal SEI can block electron transport to protect the electrolyte and passivate the electrode, while still facilitating ion diffusion[1,3]. However, in reality, electron transport across the SEI can still occur when it cracks or is insufficiently thick[18]. Currently, theoretical investigations of electron transport in the SEI are still in early stages. Multiple studies indirectly and qualitatively deduced the electron conductivity of the SEI layer by comparing electronic band structures[19-21], density of states (DOS)[19], or work function of different SEI components[17]. However, these studies cannot reflect the relationship between electron transport properties and applied bias, which is crucial given that SEIs were formed under voltage conditions[22,23]. In contrast, the non-equilibrium Green's function (NEGF) method enables direct and quantitative analysis of the electron transport properties of SEI models under bias[24-26]. So far, only a few studies have employed the NEGF method to study the electron transport properties of the SEI[24,25]. Benitez et al.[25,26] evaluated the electron transport through several SEI components using cluster models in vacuum rather than dense solid layers. Such models of SEI components may not reflect the actual working conditions of the SEI. As an improvement, Smeu et al.[24] used solid layers as structural models to investigate electron leakage through grain boundaries inside LiF on lithium–metal anodes. However, research on the electron transport properties of interfaces between different components in SEI remains lacking. Therefore, we adopted structural models similar to those used by Smeu et al. to explore the atomic-scale electron transport properties of heterogeneous interfaces between different components in the SEI on lithium metal anodes.

We employed the NEGF method to study the current density under bias voltage for LiF and $Li_2O$ layers, as well as the interfaces formed by these two components. The remaining part of this paper is organized as follows. First, through fitting the relationship between thickness and current density at 2 V, we predicted the critical thicknesses of single-crystal LiF and $Li_2O$ needed to effectively block electron transport in SEI. Then, we investigated electron transport properties of heterogeneous interfaces orthogonal to the electric-field direction by analyzing four systems with orthogonal LiF/$Li_2O$ interfaces and single-component LiF and $Li_2O$ systems. Next, we explored electron transport properties of heterogeneous interfaces parallel to the electric-field direction by studying four systems with parallel LiF/$Li_2O$ interfaces. This work reveals the contrasting effects of heterogeneous SEI interfaces orthogonal and parallel to the external electric field direction, and may help the rational design of artificial SEI and improve the performance of rechargeable batteries.

## Results and discussion

We conducted atomic-scale electron transport simulations using the SEI systems illustrated in **Fig. 1**. An SEI layer containing selected combinations of $Li_2O$ and LiF solids is placed between two semi-infinite lithium electrodes which are assumed to be in equilibrium with potentials $\mu_L$ and $\mu_R$, respectively, as shown in **Fig. 1**a. For studied systems with $LiF/Li_2O$ interfaces, the thickness of the simulated SEI layers along the electric field direction is around 1.6 nm, as shown in **Fig. 1**b. To make fair comparison between the conductivity of different SEI systems and to predict the critical thickness, simulations have first been performed on pure LiF and $Li_2O$ with various thicknesses. These SEI layers were initially built with atomic layers of LiF(001) crystal plane and/or $Li_2O$(111) crystal plane stacked along the external electric field direction, sandwiched between two lithium leads. In the other two directions orthogonal to the external field direction, the systems are assumed to be periodic, with a unit cell size of approximately 1.7 nm × 1.7 nm, as shown in **Fig. 1**c.

**Predicting critical thickness of LiF and $Li_2O$**

When the thickness reaches a critical value (i.e., the critical thickness), the tunneling probability and corresponding electric current density (i.e., the critical current density) are so low that the growth of SEI becomes negligible. To understand the influence of thickness on electron transport properties and to predict the critical thickness of single-crystal LiF and $Li_2O$, we analyzed the steady-state current density of LiF and $Li_2O$ with different thickness under bias voltages ranging from 0V to 2V. Specifically, we investigated four LiF layers with thicknesses of 0.8, 1.6, 2.4, and 3.2 nm, and six $Li_2O$ layers with thicknesses of 0.8, 1.6, 2.4, 3.2, 6.4, and 12.8 nm. Structural schematics of LiF-1.6 nm and $Li_2O$-1.6 nm systems are already shown in **Fig. 2**a-b, other structures are shown in Fig S2. **Fig. 3**a and b show I-V relationship for LiF and $Li_2O$ with different thicknesses, respectively. LiF systems consistently demonstrated current densities at least four orders of magnitude lower than equivalent $Li_2O$ system. This highlights LiF's superior effectiveness in blocking electron transport across the SEI compared to $Li_2O$, consistent with LiF being commonly recognized as a preferable candidate for an artificial SEI over $Li_2O$[27-29]. This contrasts with another NEGF study by Benitez et al.[25], where LiF's resistance is similar to $Li_2O$'s at lower bias voltages, with differences emerging only at higher bias voltages (> 4 V). The inconsistency in simulation results may stem from SEI layers being modeled as clusters in the vacuum in their study but as solid layers in ours.

Then, through fitting the relationship between thickness and current density at 2 V, we predicted the critical thickness of LiF and $Li_2O$, as shown in **Fig. 3**c and d, respectively. The logarithm of total current density ($j$) showed a robust linear relationship with SEI thickness($d$): for LiF, the relationship was described by $j/(mA\ cm^{-2})=10^{-5.21d/(nm)+12.97}$ with an $R^2$ greater than 0.99, while for $Li_2O$, it was approximately 0.98, described by $j/(mA\ cm^{-2})=10^{-0.76d/(nm)+12.67}$. The larger magnitude of the coefficient in front of $d$ for LiF (–5.21) compared to $Li_2O$ (–0.76) indicates that LiF exhibits a higher resistivity. Furthermore, taking 0.01 mA/cm² as the critical current density[30], the corresponding critical thicknesses for LiF and $Li_2O$ are around 2.9 nm and 19.0 nm, respectively. This significant difference in critical thickness again underscores LiF's superiority in blocking electron transport compared to $Li_2O$.

**Electron transport properties of heterogeneous orthogonal interfaces**

To explore the electron transport properties of heterogeneous interfaces orthogonal to the external electric field induced by bias voltage, we analyzed four systems featuring such orthogonal LiF/$Li_2O$ interfaces, alongside single-component LiF and $Li_2O$ systems for comparison. The structural configurations of these systems are depicted in **Fig. 2**a-f and Fig. S2. The four systems with LiF/Li$_2$O interfaces (depicted in **Fig. 2**c-f) feature structures containing the same number of each type of atoms. Specifically, these four simulated SEI layers consist of 4 Li/F atomic layers for LiF and 6 Li atomic layers plus 3 O atomic layers for Li$_2$O. For comparison, we also studied single-component LiF and Li$_2$O systems with thicknesses of both 0.8 nm (Fig. S2) and 1.6 nm (**Fig. 2**a-b). The 0.8 nm SEI layers consist solely of LiF (4 Li/F layers) and Li$_2$O (6 Li plus 3 O layers), matching the component counts found in the heterogeneous interface structures. For the 1.6-nm single-component systems, they contain twice the number of LiF/Li$_2$O layers while maintaining similar total thickness when compared to the four heterogeneous interface systems.

The interfaces in these heterogeneous systems are different, detailed as follows:

- The LiF/Li$_2$O system (**Fig. 2**c) includes a Li/LiF interface, a LiF/Li$_2$O interface, and a Li$_2$O/Li interface, all orthogonal to the electric field direction.
- The LiF/Li$_2$O/LiF/Li$_2$O system (**Fig. 2**d) contains a Li/LiF interface, three LiF/Li$_2$O interfaces, and a Li$_2$O/Li interface,
- The LiF/Li$_2$O/LiF system (**Fig. 2**e) includes two Li/LiF interfaces and two LiF/Li$_2$O interfaces.
- The Li$_2$O/LiF/Li$_2$O system (**Fig. 2**f) contains two LiF/Li$_2$O interfaces and two Li$_2$O/Li interfaces.

For all four systems with LiF/Li$_2$O interfaces, the structural configuration shows only slight distortion near the interfaces after optimization, indicating a favorable structural match between the LiF(001) and Li$_2$O(111) crystal planes stacked along the electric field direction.

**Fig. 4**a illustrates the steady-state current density of these systems under bias voltages ranging from 0V to 2V. First, we compared the above four interface systems

with single-component LiF and Li$_2$O systems. Neglecting the influence of orthogonal LiF/ Li$_2$O interfaces, the linear relationship established in the previous section suggests that the four systems with orthogonal interfaces would exhibit a current density of approximately $1.70\times10^8$ to $2.97\times10^8$ mA/cm$^2$ at 2V, due to the similar total thickness of LiF (~0.8 nm) and Li$_2$O (~0.8 nm) layers in these systems. However, the actual current densities are significantly lower, ranging from $4.62\times10^6$ to $7.05\times10^7$ mA/cm$^2$. This discrepancy indicates that the orthogonal heterogeneous interfaces impede electron transport.

We further compared the electron transport properties across the four configurations. Notably, the current density of the Li$_2$O/LiF/Li$_2$O/LiF system was found to be lower than that of the LiF/Li$_2$O system across the bias range from 0V to 2V. This distinction underscores the effectiveness of the orthogonal LiF/Li$_2$O interface in blocking electron transport, attributed to the former system's inclusion of three LiF/Li$_2$O interfaces compared to just one in the latter. To delve deeper into their electron transport characteristics and identify dominant energy ranges at a 2V bias, we analyzed transmission and current density spectra for both systems, as shown in Fig. S3a-b. The transmission spectrum highlights the energy ranges of the main peaks, indicative of the most conductive channels for electron transport. The current spectrum delineates energy channels contributing significantly to the overall current, with the integral area under the curves corresponding to the current at 2 V bias. Within a wide range around the Fermi energy from –0.9 eV to 0.9 eV, the LiF/Li$_2$O system exhibits a larger transmission coefficient, facilitating wider electron transport. This difference is further illustrated in the current spectrum, where the total current is primarily contributed by energy channels between –1.1 eV to 1.1 eV for both systems. Importantly, the area under the current spectrum curve is significantly larger for the LiF/Li$_2$O system compared to the LiF/Li$_2$O/LiF/Li$_2$O system, confirming the superior electron transport-blocking capability of the orthogonal LiF/Li$_2$O interface.

It should also be noted that the current density of the LiF/Li$_2$O/LiF system is approximately an order of magnitude lower than that of the Li$_2$O/LiF/Li$_2$O system. This highlights the superior electron-blocking capability of the orthogonal Li/LiF interface compared to the orthogonal Li/Li$_2$O interface, as these two systems contain the same amount of LiF and Li$_2$O layers. It is because the main structural difference between these two systems lies in the number and type of interfaces: the LiF/Li$_2$O/LiF system features two Li/LiF interfaces, whereas the Li$_2$O/LiF/Li$_2$O system features two Li/Li$_2$O interfaces. Transmission and current density spectra for both systems are provided in Fig. S3c-d. Near the bias window range, the most conductive energy channels for electron transport in the LiF/Li$_2$O/LiF system are confined narrowly from –1.4 eV to –1.3 eV, whereas those in the Li$_2$O/LiF/Li$_2$O system are broadly distributed from –1.4 eV to 1.4 eV, with significantly higher values than the LiF/Li$_2$O/LiF system spanning from –1.3 eV to 1.4 eV. Consequently, the Li$_2$O/LiF/Li$_2$O system exhibits a notably higher total probability for electron transport. This difference is further emphasized in the current spectrum shown in Fig. S3d, where the current density for Li$_2$O/LiF/Li$_2$O consistently exceeds that of LiF/Li$_2$O/LiF by more than an order of magnitude across the entire energy range. This analysis confirms the effectiveness of the orthogonal Li/LiF interface in inhibiting electron transport compared to its Li/Li$_2$O counterpart.

**Electron transport properties of parallel interfaces**

To explore electron transport properties across heterogeneous interfaces aligned parallel to the external electric field induced by bias voltage, we analysed four systems featuring parallel LiF/Li$_2$O interfaces, alongside single-component LiF and Li$_2$O systems for comparison. The structural configurations of these systems are illustrated in **Fig. 2**. The simulated SEI layers in these systems comprise varying ratios of LiF and Li$_2$O, with differing degrees of disorder. Specifically, in each periodic cell, the Li$_2$O-rich system in **Fig. 2**g consists of 4 Li/F atomic layers and 11 Li/O atomic layers; the Li$_2$O-LiF-balanced system in **Fig. 2**h consists of 6 Li/F atomic layers and 8 Li/O atomic layers; and the LiF-rich system in **Fig. 2**i consists of 8 Li/F atomic layers and 5 Li/O atomic layers. **Fig. 2**j shows the twice-interface system with the same total number of atomic layers as the balanced system, but with interleaved atomic layers resulting in twice as many interfaces parallel to the electric field direction in each unit cell. In each periodic cell, the 1.6 nm single-component LiF system in **Fig. 2**a contains 12 Li/F atomic layers aligned parallel to the electric field, while the 1.6 nm single- component Li$_2$O system in **Fig. 2**b contains 18 Li/O atomic layers similarly aligned.

In each periodic cell, the interfaces in these heterogeneous systems vary as detailed below:

- The Li$_2$O-rich system (**Fig. 2**g) features two LiF/Li$_2$O interfaces parallel to the electric field direction, with Li$_2$O thickness comprising roughly two thirds of the total thickness.
- The Li$_2$O-LiF-balanced system (**Fig. 2**h) includes two LiF/Li$_2$O interfaces, with approximately equal thicknesses for LiF and Li$_2$O.
- The LiF-rich system (**Fig. 2**i) contains two LiF/Li$_2$O interfaces, with Li$_2$O thickness comprising roughly one third of the total thickness.
- The twice-interface system (**Fig. 2**j) incorporates four LiF/Li$_2$O interfaces, each with roughly equal distances (approximately one fourth of the total thickness) between adjacent interfaces. The reduced thickness of each layer and the increased number of interfaces lead to a much more disordered structure compared to the other three.

These structures exhibit greater distortion near the LiF/Li$_2$O interfaces compared to the orthogonal cases, indicating a less ideal match between the LiF(001) and Li$_2$O(111) layers stacked parallel to the external field direction, possibly due to the formation of Li-LiF-Li$_2$O triple phase boundaries.

**Fig. 4**b illustrates the steady-state current density of these systems under bias voltages ranging from 0 V to 2 V. First, we compared four systems featuring LiF/Li$_2$O interfaces with single-component LiF and Li$_2$O systems. The current density of four systems with LiF/Li$_2$O interfaces is substantially lower than that of Li$_2$O at 1.6 nm, but significantly higher than that of LiF at 1.6 nm. This indicates that electron transport in structures with parallel interfaces is predominantly determined by the component with higher electron transport capability. In Li$_2$O-rich, Li$_2$O-LiF-balanced, and LiF-rich systems, the current density decreases with reduced Li$_2$O content, reflecting Li$_2$O's superior conductivity. As shown in Fig. S4a, transmission spectra at 2 V reveals a conductive energy range from –1.4 eV to –0.6 eV for pure Li$_2$O, Li$_2$O-rich, and Li$_2$O-LiF-balanced systems. LiF-rich system exhibits the most conductive energy range narrowing to –1.4 eV to –1.3 eV, with transmission levels in this range lower by approximately an order of magnitude compared to the other systems. Furthermore, the transmission spectra in Fig. S4c confirmed that the pure LiF system has significantly lower transmission levels compared to all other systems, with values at least four orders of magnitude smaller.

Surprisingly, above 1.5 V, the Li$_2$O-rich system exhibits current densities comparable to or higher than pure Li$_2$O, indicating enhanced electron transport through the SEI layer via parallel interfaces, contrary to findings from orthogonal systems. As shown in the current spectrum in Fig. S4b, the Li$_2$O-rich system is more conductive than the pure Li$_2$O from –0.6 to –0.9 eV, which constitutes the main contribution to the electric current. This explains the higher current density of the Li$_2$O-rich system and highlights the electron transport enhancement by parallel interfaces. This observation is also consistent with the NEGF study by Smeu et. al.[24], where a similarly oriented boundary between LiF grains of different crystal orientations enhanced the leakage current.

It should also be noted that the current density of the twice-interface system is more than an order of magnitude lower than that of the $Li_2O$-LiF-balanced system, despite containing the same total amount of LiF and $Li_2O$. The increase in interface density leads to a higher degree of structural disorder, which significantly disrupts electron transport. This is evident when comparing the two systems: the $Li_2O$-LiF-balanced system, with its two LiF/$Li_2O$ interfaces and four triple phase boundaries per unit cell, contrasts sharply with the twice-interface system, which has four interfaces and eight triple phase boundaries. As depicted in the transmission spectrum in Fig. S4e, for the $Li_2O$-LiF-balanced system, the most conductive energy channels for electron transport span from –1.4 eV to –0.5 eV, narrowing to –1.4 eV to –0.8 eV as the number of interfaces is doubled, accompanied by a noticeable decrease in transmission. This difference is further illustrated in the current spectrum in Fig. S4f, where the twice-interface system shows a four-fold decrease. This analysis confirms the disruptive role of structural disorder in electron transport across the entire conductive energy range.

The insights gained from our study provide a profound understanding of the phenomena occurring in rechargeable batteries from the perspective of electron transport. We have found that orthogonal interfaces and structural disorders act as impediments to electron transport, thereby enhancing electron-blockage and passivation effects. Conversely, parallel interfaces exhibit a different behavior. This implies that in the design of artificial SEIs, thinner layers with a higher prevalence of orthogonal interfaces and structural disorders can deliver comparable passivation benefits as thicker layers with a greater presence of parallel interfaces. Furthermore, our findings suggest that during the cycling of rechargeable batteries, dendrite growth may be slower at orthogonal interfaces and locations with structural disorders compared to regions with parallel interfaces.

## Conclusions

In summary, we employed the NEGF method to theoretically investigate atomic-scale electron transport properties under bias for LiF/Li$_2$O interfaces and single-component atomic layers of these two components. We predicted that the critical thicknesses for effectively blocking electron transport in single-crystal LiF and Li$_2$O would be 2.9 nm and 19.0 nm, respectively. Furthermore, we found that LiF/Li$_2$O interfaces orthogonal to the electric-field direction impede electron transport. Conversely, LiF/Li$_2$O interfaces parallel to the external electric-field direction can enhance electron transport but structural disorder induced by densely distributed interfaces severely interferes with electron transport. This study provides valuable insights for rational design of effective artificial SEI layers, and may further lay the foundation for improving the performance of rechargeable batteries.

## Methodology

**Non-equilibrium Green's function for studying electron transport**

The electron transport properties of the systems are studied with the non-equilibrium Green's function (NEGF) method. The key construction in NEGF is the retarded Green's function of the reduced system comprising the central region, and it can be calculated using the following formula,

$$G(E) = [ES - H - \Sigma_\text{L}(E) - \Sigma_\text{R}(E)^{-1}] \quad (1)$$

where $S$ and $H$ stand for the overlap matrix and the Hamiltonian of the reduced system. The $\Sigma_\text{L}$ and $\Sigma_\text{R}$ are the self-energies matrices for the left and right electrodes, respectively. They are complex energy-dependent matrices and account for the influence of the electrodes on the central region. The real parts of the self-energies account for the shift of the energy levels. The imaginary parts, $\Gamma_\text{L,R} = i(\Sigma_\text{L,R} - \Sigma_\text{L,R}^\dagger)$, are known as the line-width function and account for the broadening. The self-energies are evaluated iteratively from the electrode and device Hamiltonians. The expression for the electron density matrix is as follows,

$$\rho = \frac{1}{2\pi}\int_{-\infty}^{\infty}[f(E,\mu_\text{L})G\Gamma_\text{L}G^\dagger + f(E,\mu_\text{R})G\Gamma_\text{R}G^\dagger]dE \tag{2}$$

where the Fermi distribution functions $f(E,\mu_\text{L})$ and $f(E,\mu_\text{R})$ give the occupancy of the electron orbital at energy $E$ for an electrode at equilibrium with electrochemical potential $\mu_\text{L,R}$. The difference between the two electrochemical potentials gives the bias voltage $eV_\text{b} = \mu_\text{L} - \mu_\text{R}$. The electron density matrix calculated above is used to update the local potential distribution, which is in turn used to update the Hamiltonian and the retarded Green's function. This process is proceeded iteratively until self-consistency is achieved. The steady-state transmission function is then evaluated using the converged Green's function as

$$T(E,V_\text{b}) = \text{Tr}(\Gamma_\text{L}G\Gamma_\text{R}G^\dagger) \tag{3}$$

which describes the transmission probability of an electron to transport between the electrodes via a channel with energy $E$ in the central region. The product of the transmission function and the difference of the Fermi functions of the two electrodes $\frac{dI}{dE} = \frac{2e}{h}T(E,V_\text{b})[f(E,\mu_\text{L}) - f(E,\mu_\text{R})]$ is known as the current spectrum. It indicates the contribution to the total current from the specific energy channel. Its integration gives back the Landauer-Büttiker formula for total current[31],

$$I(V_b) = \frac{2e}{h} \int_{-\infty}^{\infty} T(E, V_b)[f(E, \mu_L) - f(E, \mu_R)]dE \qquad (4)$$

The transmission per unit area, current spectrum per unit area, and current density ($j$) are obtained through dividing the area of the interfaces.

**Computational methods**

Our simulation integrates the DFT method for structural optimization with the NEGF-DFTB method for electronic transport property calculations. The DFT method provides relatively precise descriptions of SEI layer structures, whereas the DFTB method, known for its efficiency, enables NEGF simulations of SEI layers with significantly greater thicknesses[32,33]. The method of building structures for these two steps is illustrated in Fig. S1, and VESTA is employed for visualization of all structures.

The Vienna Ab initio Simulation Package (VASP, version 5.4.4) is used for structural relaxation. The Perdew–Burke–Ernzerhof[34] (PBE) approximation is used for exchange-correlation energy. The projector augmented wave (PAW) method handles electron interactions using a plane-wave basis expanded to 500 eV[35-37]. The Monkhorst–Pack scheme samples the 1 × 1 × 1 Brillouin zone, and Gaussian smearing (0.05 eV) is used as the occupation method. The electronic self-consistent convergence criterion is $10^{-5}$ eV.

The electron transport calculations are performed with LODESTAR[38]. DFTB parameters are generated using our own software by fitting electronic structures of various solid systems containing Li, F, and O atoms against DFT results at PBE level [39,40]. The structure for transport simulation is constructed by replicating two layers of fixed lithium atoms from the optimized structure on both sides. Initially, DFTB Mulliken charges for the electrodes are derived from simulations on infinite equilibrium systems. These charges are subsequently used in a self-consistent process to solve the Poisson equation and NEGF equations for the entire system under specific bias conditions, yielding Mulliken charges for the device region. Finally, electronic transport properties, including transmission coefficients and current spectra, are computed through integration following the NEGF scheme.

## Data availability

The data that support the plots within this paper and other findings of this study are available from the corresponding author upon reasonable request.

## References

1    Xu, K. Electrolytes and Interphases in Li-Ion Batteries and Beyond. *Chemical Reviews* **114**, 11503-11618 (2014). https://doi.org/10.1021/cr500003w

2    Peled, E. The Electrochemical Behavior of Alkali and Alkaline Earth Metals in Nonaqueous Battery Systems—The Solid Electrolyte Interphase Model. *J. Electrochem. Soc.* **126**, 2047 (1979). https://doi.org/10.1149/1.2128859

3    Wang, A., Kadam, S., Li, H., Shi, S. & Qi, Y. Review on modeling of the anode solid electrolyte interphase (SEI) for lithium-ion batteries. *npj Computational Materials* **4**, 15 (2018). https://doi.org/10.1038/s41524-018-0064-0

4    Xu, K. Interfaces and interphases in batteries. *Journal of Power Sources* **559**, 232652 (2023). https://doi.org/https://doi.org/10.1016/j.jpowsour.2023.232652

5	Lin, D., Liu, Y. & Cui, Y. Reviving the lithium metal anode for high-energy batteries. *Nature Nanotechnology* **12**, 194-206 (2017). https://doi.org/10.1038/nnano.2017.16

6	Winter, M. The Solid Electrolyte Interphase – The Most Important and the Least Understood Solid Electrolyte in Rechargeable Li Batteries.   **223**, 1395-1406 (2009). https://doi.org/doi:10.1524/zpch.2009.6086

7	He, X. *et al.* The passivity of lithium electrodes in liquid electrolytes for secondary batteries. *Nature Reviews Materials* **6**, 1036-1052 (2021). https://doi.org/10.1038/s41578-021-00345-5

8	Peled, E. & Menkin, S. Review—SEI: Past, Present and Future. *J. Electrochem. Soc.* **164**, A1703 (2017). https://doi.org/10.1149/2.1441707jes

9	Liu, W., Liu, P. & Mitlin, D. Review of Emerging Concepts in SEI Analysis and Artificial SEI Membranes for Lithium, Sodium, and Potassium Metal Battery Anodes. *Advanced Energy Materials* **10**, 2002297 (2020). https://doi.org/https://doi.org/10.1002/aenm.202002297

10	Fedorov, R. G., Maletti, S., Heubner, C., Michaelis, A. & Ein-Eli, Y. Molecular Engineering Approaches to Fabricate Artificial Solid-Electrolyte Interphases on Anodes for Li-Ion Batteries: A Critical Review. *Advanced Energy Materials* **11**, 2101173 (2021). https://doi.org/https://doi.org/10.1002/aenm.202101173

11	Adenusi, H., Chass, G. A., Passerini, S., Tian, K. V. & Chen, G. Lithium Batteries and the Solid Electrolyte Interphase (SEI)—Progress and Outlook. *Advanced Energy Materials* **13**, 2203307 (2023). https://doi.org/https://doi.org/10.1002/aenm.202203307

12	Heiskanen, S. K., Kim, J. & Lucht, B. L. Generation and Evolution of the Solid Electrolyte Interphase of Lithium-Ion Batteries. *Joule* **3**, 2322-2333 (2019). https://doi.org/https://doi.org/10.1016/j.joule.2019.08.018

13	Aurbach, D. *et al.* An analysis of rechargeable lithium-ion batteries after prolonged cycling. *Electrochimica Acta* **47**, 1899-1911 (2002). https://doi.org/https://doi.org/10.1016/S0013-4686(02)00013-0

14	Ramasubramanian, A. *et al.* Stability of Solid-Electrolyte Interphase (SEI) on the Lithium Metal Surface in Lithium Metal Batteries (LMBs). *ACS Applied Energy Materials* **3**, 10560-10567 (2020).

https://doi.org/10.1021/acsaem.0c01605

15   Ramasubramanian, A. *et al.* Lithium Diffusion Mechanism through Solid–Electrolyte Interphase in Rechargeable Lithium Batteries. *The Journal of Physical Chemistry C* **123**, 10237-10245 (2019). https://doi.org/10.1021/acs.jpcc.9b00436

16   Tian, H.-K., Liu, Z., Ji, Y., Chen, L.-Q. & Qi, Y. Interfacial Electronic Properties Dictate Li Dendrite Growth in Solid Electrolytes. *Chemistry of Materials* **31**, 7351-7359 (2019). https://doi.org/10.1021/acs.chemmater.9b01967

17   Lin, Y.-X. *et al.* Connecting the irreversible capacity loss in Li-ion batteries with the electronic insulating properties of solid electrolyte interphase (SEI) components. *Journal of Power Sources* **309**, 221-230 (2016). https://doi.org/https://doi.org/10.1016/j.jpowsour.2016.01.078

18   Köbbing, L., Latz, A. & Horstmann, B. Growth of the solid-electrolyte interphase: Electron diffusion versus solvent diffusion. *Journal of Power Sources* **561**, 232651 (2023). https://doi.org/https://doi.org/10.1016/j.jpowsour.2023.232651

19   Luo, L. *et al.* Tuning the electron transport behavior at Li/LATP interface for enhanced cyclability of solid-state Li batteries. *Nano Research* **16**, 1634-1641 (2023). https://doi.org/10.1007/s12274-022-5136-2

20   Liu, Z. *et al.* Interfacial Study on Solid Electrolyte Interphase at Li Metal Anode: Implication for Li Dendrite Growth. *J. Electrochem. Soc.* **163**, A592 (2016). https://doi.org/10.1149/2.0151605jes

21   Wang, E. *et al.* Mitigating Electron Leakage of Solid Electrolyte Interface for Stable Sodium-Ion Batteries. *Angewandte Chemie International Edition* **62**, e202216354 (2023). https://doi.org/https://doi.org/10.1002/anie.202216354

22   Xu, Y. *et al.* Direct in situ measurements of electrical properties of solid–electrolyte interphase on lithium metal anodes. *Nature Energy* **8**, 1345-1354 (2023). https://doi.org/10.1038/s41560-023-01361-1

23   Qi, Y. Measuring is believing. *Nature Energy* **8**, 1307-1308 (2023). https://doi.org/10.1038/s41560-023-01371-z

24   Smeu, M. & Leung, K. Electron leakage through heterogeneous LiF

on lithium–metal battery anodes. *Physical Chemistry Chemical Physics* **23**, 3214-3218 (2021). https://doi.org/10.1039/D0CP06310J

25   Benitez, L., Cristancho, D., Seminario, J. M., Martinez de la Hoz, J. M. & Balbuena, P. B. Electron transfer through solid-electrolyte-interphase layers formed on Si anodes of Li-ion batteries. *Electrochimica Acta* **140**, 250-257 (2014). https://doi.org/https://doi.org/10.1016/j.electacta.2014.05.018

26   Benitez, L. & Seminario, J. M. Electron Transport and Electrolyte Reduction in the Solid-Electrolyte Interphase of Rechargeable Lithium Ion Batteries with Silicon Anodes. *The Journal of Physical Chemistry C* **120**, 17978-17988 (2016). https://doi.org/10.1021/acs.jpcc.6b06446

27   Jian Hu, X. *et al.* Artificial LiF-Rich Interface Enabled by In situ Electrochemical Fluorination for Stable Lithium-Metal Batteries. *Angewandte Chemie International Edition* **63**, e202319600 (2024). https://doi.org/https://doi.org/10.1002/anie.202319600

28   Yu, T. *et al.* Spatially Confined LiF Nanoparticles in an Aligned Polymer Matrix as the Artificial SEI Layer for Lithium Metal Anodes. *Nano Letters* **23**, 276-282 (2023). https://doi.org/10.1021/acs.nanolett.2c04242

29   Zhang, L., Zhang, K., Shi, Z. & Zhang, S. LiF as an Artificial SEI Layer to Enhance the High-Temperature Cycle Performance of Li4Ti5O12. *Langmuir* **33**, 11164-11169 (2017). https://doi.org/10.1021/acs.langmuir.7b02031

30   Li, D. *et al.* Modeling the SEI-Formation on Graphite Electrodes in LiFePO$_4$ Batteries. *J. Electrochem. Soc.* **162**, A858 (2015). https://doi.org/10.1149/2.0161506jes

31   Datta, S. *Electronic transport in mesoscopic systems*.   (Cambridge University Press, 1995).

32   Elstner, M. *et al.* Self-consistent-charge density-functional tight-binding method for simulations of complex materials properties. *Physical Review B* **58**, 7260-7268 (1998). https://doi.org/10.1103/PhysRevB.58.7260

33   Elstner, M., Frauenheim, T. & Suhai, S. An approximate DFT method for QM/MM simulations of biological structures and processes. *Journal of Molecular Structure: THEOCHEM* **632**, 29-41 (2003). https://doi.org/https://doi.org/10.1016/S0166-1280(03)00286-0

34  Perdew, J. P., Burke, K. & Ernzerhof, M. Generalized Gradient Approximation Made Simple. *Physical Review Letters* **77**, 3865-3868 (1996). https://doi.org/10.1103/PhysRevLett.77.3865

35  Kresse, G. & Furthmüller, J. Efficiency of ab-initio total energy calculations for metals and semiconductors using a plane-wave basis set. *Computational Materials Science* **6**, 15-50 (1996). https://doi.org/https://doi.org/10.1016/0927-0256(96)00008-0

36  Kresse, G. & Joubert, D. From ultrasoft pseudopotentials to the projector augmented-wave method. *Physical Review B* **59**, 1758-1775 (1999). https://doi.org/10.1103/PhysRevB.59.1758

37  Kresse, G. & Furthmüller, J. Efficient iterative schemes for ab initio total-energy calculations using a plane-wave basis set. *Physical Review B* **54**, 11169-11186 (1996). https://doi.org/10.1103/PhysRevB.54.11169

38  Chen, S., Kwok, Y. & Chen, G. Time-Dependent Density Functional Theory for Open Systems and Its Applications. *Accounts of Chemical Research* **51**, 385-393 (2018). https://doi.org/10.1021/acs.accounts.7b00382

39  Gao, R. *et al.* Self-Consistent-Charge Density-Functional Tight-Binding Parameters for Modeling an All-Solid-State Lithium Battery. *Journal of Chemical Theory and Computation* (2023). https://doi.org/10.1021/acs.jctc.2c01115

40  Porezag, D., Frauenheim, T., Köhler, T., Seifert, G. & Kaschner, R. Construction of tight-binding-like potentials on the basis of density-functional theory: Application to carbon. *Physical Review B* **51**, 12947-12957 (1995). https://doi.org/10.1103/PhysRevB.51.12947

## Acknowledgements

GHC acknowledges financial support by the General Research Fund (Grant No. 17309620) and Research Grants Council (RGC: T23-713/22-R). GHC acknowledges support from the Hong Kong Quantum AI Lab, AIR@InnoHK of the Hong Kong Government.

## Competing interests

The authors declare no competing interests.

# Figures

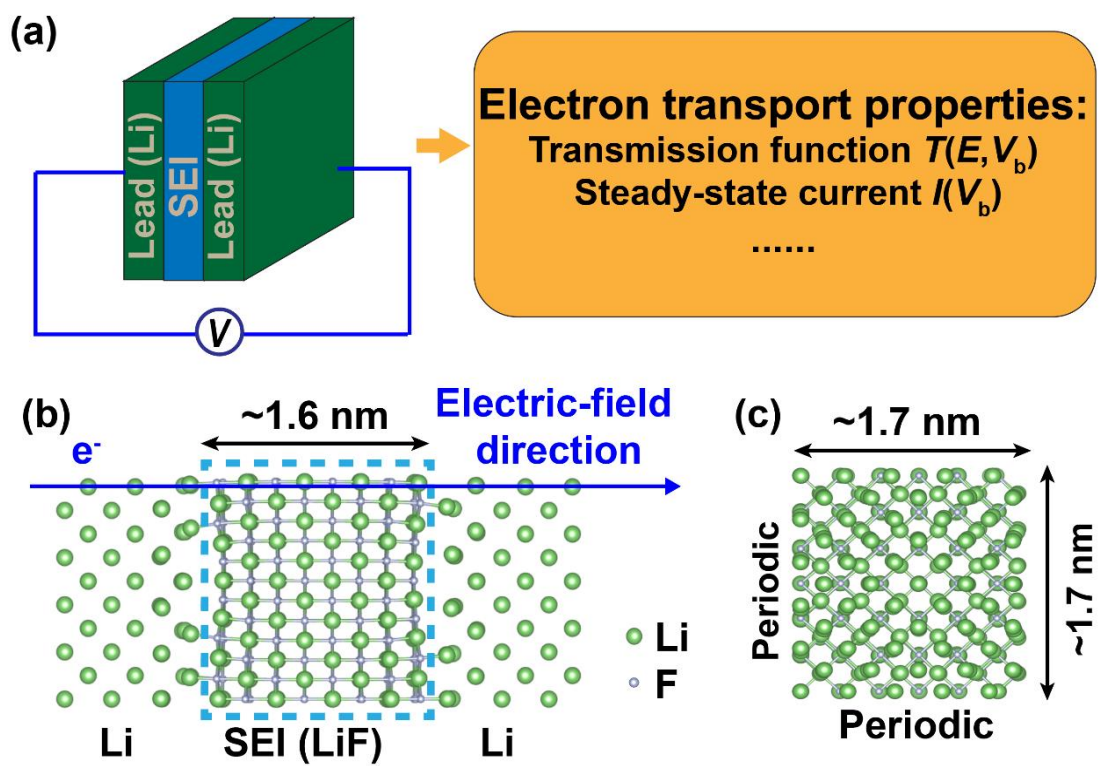

**Fig. 1.** Schematics of the SEI systems for simulating electron transport properties. **a**, Schematic of the whole circuit. The SEI layer (blue) is placed between two lithium electrodes (green), where the external bias voltage is applied. **b**, Schematic of atomic configuration of SEI systems for NEGF simulation, demonstrated by the side-view of LiF-1.6 nm system. The SEI layer is indicated by the dashed blue box. For all studied systems with LiF/Li$_2$O interfaces, the thicknesses of SEI layers are around 1.6 nm. The direction of external electric field induced by bias voltage goes from one electrode to the other, as indicated by the blue arrow. **c**, Cross-section of a unit cell of the system in (**b**), for demonstrating the periodicity and size of studied systems. All studied systems are periodic in directions orthogonal to the electric field direction, with a unit cell length around 1.7 nm.

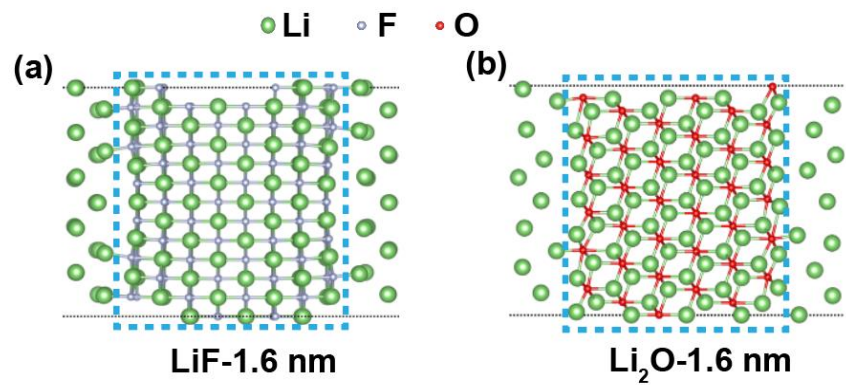
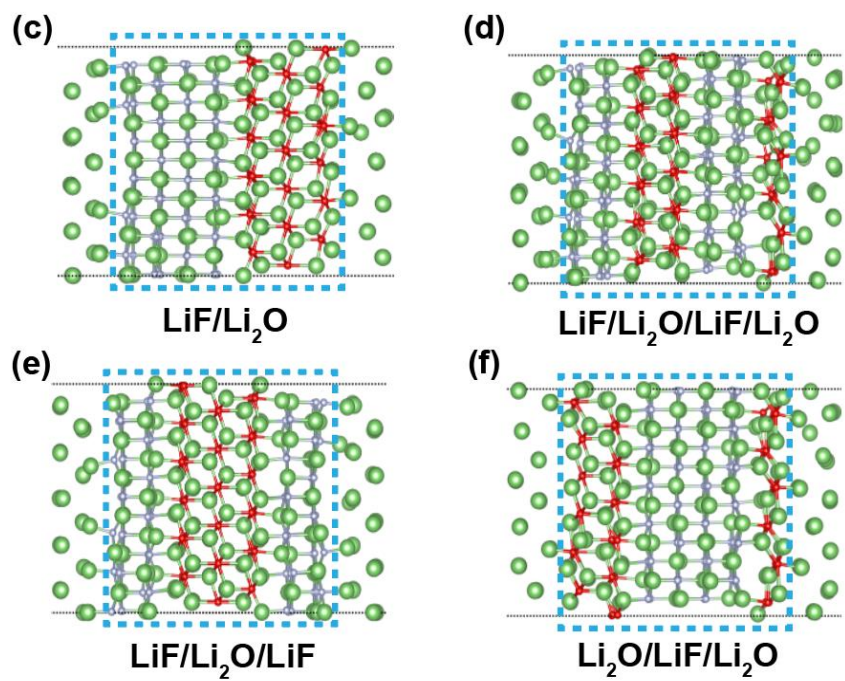
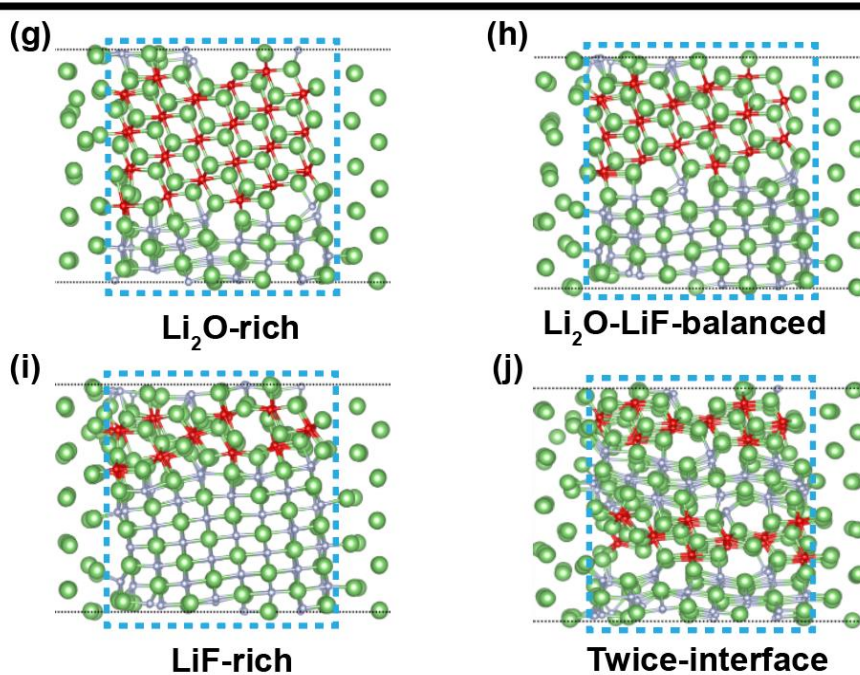

**Fig. 2.** Structures for studied system with SEI layers around 1.6 nm. **a**, the LiF-1.6 nm system. **b**, the $Li_2O$-1.6 nm system. **c-f** show four systems with heterogeneous SEI interfaces orthogonal to the electric field direction, while **g-j** show four systems with parallel-oriented SEI interfaces. **c**, the LiF/$Li_2O$ system, containing a LiF layer and a $Li_2O$ layer of roughly equal thickness. **d**, the LiF/$Li_2O$/LiF/$Li_2O$ system, containing the same amount for each component as in (**c**), but twice as many heterogeneous interfaces. **e**, the LiF/$Li_2O$/LiF system, with two LiF/Li interfaces. **f**, the $Li_2O$/LiF/$Li_2O$ system, with two $Li_2O$/Li interfaces. **g**, the $Li_2O$-rich system, comprising a thicker layer of $Li_2O$ than LiF. **h**, the $Li_2O$-LiF-balanced system, comprising a layer of $Li_2O$ and a layer of LiF of comparable thicknesses. **i**, the LiF-rich system, comprising a thicker layer of LiF than $Li_2O$. **j**, the twice-interface system, comprising the same amount of $Li_2O$ and LiF as (**h**) but with denser interface population. The blue dashed box in each structure indicates the SEI layer with thickness around 1.6nm.

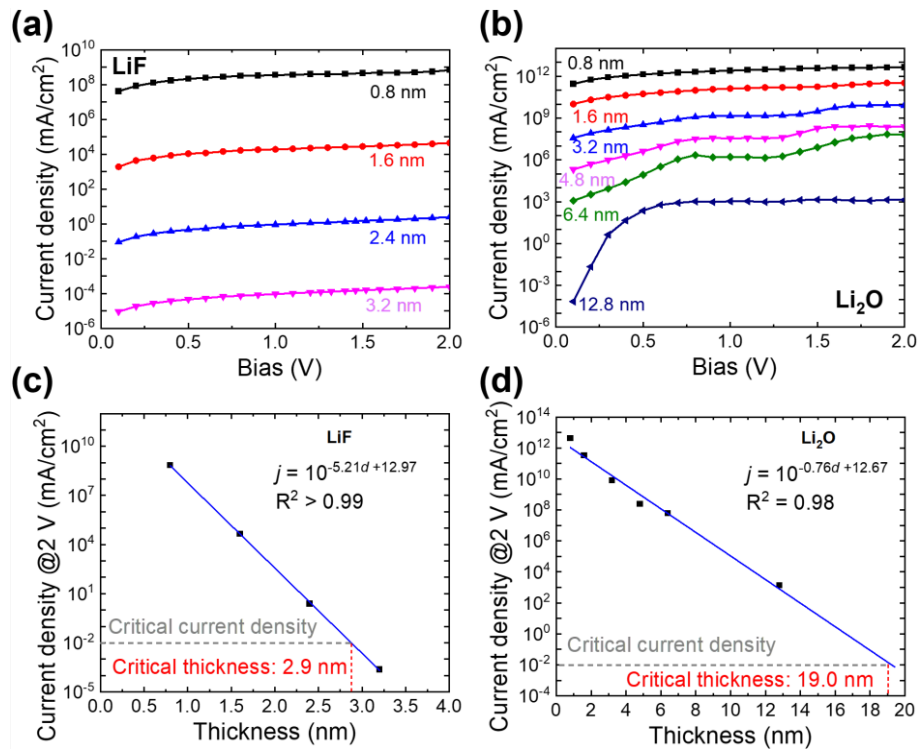

**Fig. 3.** Relation between electronic current density and SEI layer thickness, highlighting the critical thicknesses. **a**, I-V curves for pure LiF SEI layer with thicknesses 0.8, 1.6, 2.4, and 3.2 nm. **b**, I-V curves for pure $Li_2O$ SEI layer with thicknesses 0.8, 1.6, 2.4, 3.2, 6.4, and 12.8 nm. **c**, correlation analysis of current density at 2V against thickness for LiF systems, showing the critical thickness 2.9 nm corresponding to critical current density 0.01 mA/cm$^2$. **d**, correlation analysis of current density at 2 V against thickness for $Li_2O$ systems, showing the critical thickness 19.0 nm corresponding to critical current density 0.01 mA/cm$^2$. In (**c**, **d**), the variables *d* and *j* represent the thickness (nm) of the studied single component and the current density (mA/cm$^2$) of the investigated systems at a voltage of 2 V, respectively.

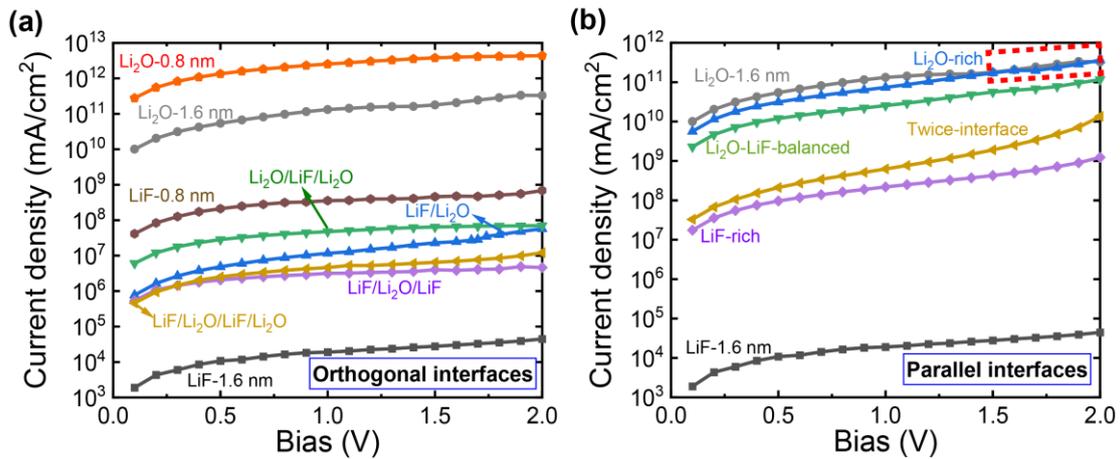

**Fig. 4.** I-V curves for studied systems with heterogeneous SEI interfaces under a bias range of 0~2V. **a** and (**b**) show orthogonal and parallel-oriented interfaces systems, respectively, with single-component systems included for comparison. (**a**) indicates that heterogeneous interfaces orthogonal to the electric field direction leads to a decreased current density. (**b**) shows the lower current density of the twice-interface system (shown in Fig. 2j) than $Li_2O$-LiF-balanced system the system (shown in Fig. 2h) as denser interfaces are included in the former system. The grey dashed box highlights the crossing of current between the Li$_2$O-rich system (shown in Fig. 2g) and Li$_2$O-1.6 nm system at high bias voltage.

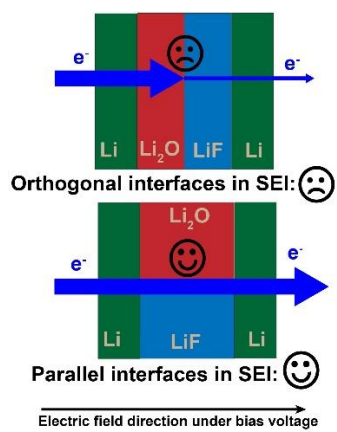

Graphical abstract